\documentclass[runningheads]{llncs}

\usepackage{graphicx}
\usepackage{xcolor}
\usepackage{amsmath}
\usepackage{amssymb}
\usepackage{booktabs}
\usepackage{comment}
\usepackage[misc]{ifsym}

\usepackage{hyperref}

\usepackage{xspace}
\newcommand{\todocite}[1]{\textcolor{red}{[cite!\ifthenelse{\equal{#1}{}}{}{ #1}]}\xspace}
\newcommand{\todo}[1]{\textcolor{red}{(todo\emph{\ifthenelse{\equal{#1}{}}{}{ #1}!)}}\xspace}

\usepackage{color}

\definecolor{ben}{rgb}{0.9,0.,0.5}

\definecolor{vincent}{rgb}{0.,0.5,0.5}

\definecolor{alex}{rgb}{0.,0.,0.8}

\definecolor{hayoung}{rgb}{0.5,0.2,0.8}

\definecolor{mahdi}{rgb}{0.1,0.7,0.9}

\usepackage{siunitx}
\DeclareSIUnit{\nothing}{\relax}

\usepackage{multirow}

\makeatletter
\newcommand{\printfnsymbol}[1]{%
  \textsuperscript{\@fnsymbol{#1}}%
}
\makeatother
\begin{document}

\title{On the Localization of Ultrasound Image Slices within Point Distribution Models}
\titlerunning{Localizing Ultrasound Image Slices Within PDMs}
\author{
Lennart Bastian\thanks{denotes equal contribution. \qquad \qquad \Letter: \email{lennart.bastian@tum.de}} \textsuperscript{\Letter} \and 
Vincent Bürgin\printfnsymbol{1} \and
Ha Young Kim\printfnsymbol{1} \and \\
Alexander Baumann \and
Benjamin Busam \and
Mahdi Saleh \and
Nassir Navab
}

\authorrunning{L. Bastian et al.}
\institute{Computer Aided Medical Procedures, Technical University of Munich, Germany}
\maketitle              %
\begin{abstract}
Thyroid disorders are most commonly diagnosed using high-resolution Ultrasound (US).
Longitudinal nodule tracking is a pivotal diagnostic protocol for monitoring changes in pathological thyroid morphology.
This task, however, imposes a substantial cognitive load on clinicians due to the inherent challenge of maintaining a mental 3D reconstruction of the organ.
We thus present a framework for automated US image slice localization within a 3D shape representation to ease how such sonographic diagnoses are carried out.
Our proposed method learns a common latent embedding space between US image patches and the 3D surface of an individual's thyroid shape, or a statistical aggregation in the form of a statistical shape model (SSM), via contrastive metric learning.
Using cross-modality registration and Procrustes analysis, we leverage features from our model to register US slices to a 3D mesh representation of the thyroid shape.
We demonstrate that our multi-modal registration framework can localize images on the 3D surface topology of a patient-specific organ and the mean shape of an SSM.
Experimental results indicate slice positions can be predicted within an average of 1.2 mm of the ground-truth slice location on the patient-specific 3D anatomy and 4.6 mm on the SSM, exemplifying its usefulness for slice localization during sonographic acquisitions. Code is publically available: \href{https://github.com/vuenc/slice-to-shape}{https://github.com/vuenc/slice-to-shape}

\end{abstract}

\section{Introduction}
\label{sec:intro}

\begin{figure}[!ht]
    \centering
    \includegraphics[width=1.0\textwidth]{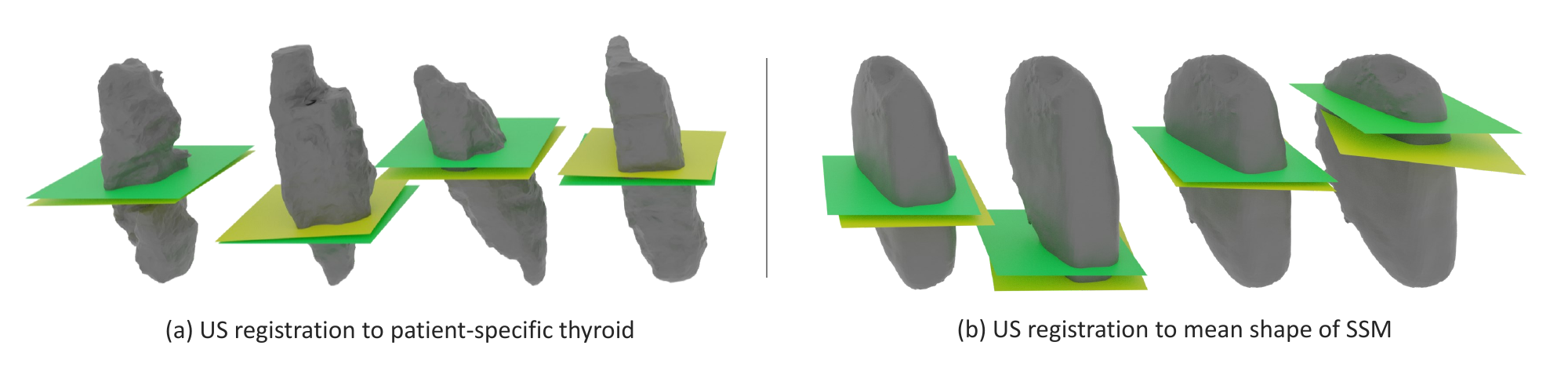}
    \caption{
    We propose the task of slice-to-shape registration. Our method successfully localizes ultrasound images to 3D thyroid shapes from the same individual (a) and the mean of a statistical shape model (b). 
    The ground truth image plane is depicted in green, and the prediction in yellow. 
    }
    \label{fig:example_thyroids}
\end{figure}

High-resolution Ultrasound (US) has detected the presence of thyroid nodules in up to 69\% of scans in randomly selected individuals~\cite{haugen2015AmericanThyroid2016}. 
While only 7-15\% of cases develop into malignant tumors, periodic screening is an essential prophylactic measure in the early diagnosis and treatment of a debilitating disease. 
B-mode US has been identified as the primary tool for diagnosing malignant thyroid nodules for its ease of use, lack of harmful ionizing radiation \cite{azizi3DUltrasoundThyroid2021}, and exceptional soft-tissue resolution \cite{chengUse3DUltrasound2022}.
A rise in the prevalence of thyroid cancer over the past decades has been mainly attributed to an increase in early detection with the help of more frequent US screenings \cite{haugen2015AmericanThyroid2016}.
However, US is highly operator-dependent and requires significant training to generate clear and accurate images. 
Furthermore, US images are abundant with noise and artifacts induced by physical properties such as phase aberrations and attenuation, which can introduce uncertainty and yield inconsistent diagnoses in thyroid nodule classification \cite{chengUse3DUltrasound2022} and thyroid volume \cite{kronkeTracked3DUltrasound2022} estimation.
Methods that could improve thyroid scanning and US image quality are therefore highly sought after to increase early detection rates of thyroid cancer worldwide.

Observing a thyroid nodule's evolution is a critical diagnostic protocol \cite{chengUse3DUltrasound2022}.
To understand how a nodule changes over time and if it is potentially developing into a malignant tumor, clinicians meticulously match the thyroid morphology to a previous US acquisition.
Such a procedure requires significant dexterity, elaborate training, and a cognitive 3D reconstruction while simultaneously conducting a complex medical evaluation.
Providing automated support for this procedure could not only alleviate the cognitive burden on clinicians but also reduce costs by enabling less experienced individuals to conduct these scans semi-autonomously. 
Furthermore, this technological aid could extend potentially life-saving diagnoses to remote communities that lack specialized medical expertise \cite{naceri2022tactile,zhangClinicalApplication5Gbased2022,chai2022successful}.
A fundamental challenge in automating such a procedure is intra-organ localization.

To address this problem generally, we therefore propose a framework for slice-to-shape registration.
Existing 2D-3D registration methods in the medical domain either register image slices to individual 3D volumes \cite{songCrossModalAttentionMRI2021,markovaGlobalMultimodal2D2022a} or aggregated volumes in the form of a medical atlas \cite{yeungAdaptive3DLocalization2022}.
While image-atlas-based methods have been demonstrated effective for navigational support, deformable image registration is ill-conditioned and difficult to regularize \cite{fu2020deep}. 
Furthermore, registration inaccuracies can yield unrepresentative voxel intensities in the atlas, adversely affecting downstream slice-to-atlas registration.
We propose the task of slice-to-shape registration, by registering US slices to a 3D shape representation either directly obtained from an individual's organ segmentation contour, or aggregated shapes in the form of a statistical shape model (SSM) for a more general localization approach which does not require pre-operative acquisitions.

The medical imaging literature has not explored the registration of medical images to 3D organ shapes, particularly for statistical point distributions (see Figure \ref{fig:example_thyroids}).
We thus propose a self-supervised metric learning pipeline to enable matching and registration across US images and a 3D mesh representation of an organ.
We leverage unsupervised correspondence estimation to generate a point distribution model (PDM) \cite{bastian2023s3m} and use these correspondences to map image patches to a corresponding location on SSM shapes during training.
Patch features are extracted using separate deep neural networks, and their cross-modal representations are used to localize a US query slice inside the SSM. 
Despite having limited supervision and learning on a geometric surface representation, our pipeline for partial thyroid registration successfully localizes US slices from unseen subjects in an SSM.

Our main contributions can be summarized as follows:
\begin{itemize}
\item We propose the task of slice-to-shape registration in medical imaging for 3D organ shapes and SSMs.
\item Our slice-to-shape correspondence pipeline enables registration of ultrasound slices to 3D thyroid shapes through multi-modal contrastive metric learning.
\item We evaluate the capabilities of our model for US slice localization on patient-specific 3D thyroid meshes and SSMs, demonstrating for the first time that 2D US images can be localized within a geometric statistical distribution without prior patient-specific acquisitions.
\end{itemize}

\section{Related Work}

\subsection{SSMs and Image Atlases in Medical Imaging}

Statistical Shape Models (SSMs) and Image Atlases have distinct yet interconnected roles in medical imaging. 
Since the early 1990s, SSMs have been widely employed for their efficient encapsulation of shape variations and usefulness in enhancing the robustness of segmentation techniques \cite{heimannStatisticalShapeModels2009}.
Both methodologies have found utility in diverse applications such as segmentation \cite{raju2022deep}, registration \cite{ellingsenRobustDeformableImage2010,berendsen2013free}, shape classification \cite{grassi2021statistical,ludke2022landmark}, and image augmentation \cite{tang2019augmentation,uzunova2022systematic}. More recently, SSMs have been used as priors to improve the robustness of medical image segmentation in deep neural networks \cite{raju2022deep}, enhance myocardial motion tracking \cite{hu2022deep}, and facilitate the segmentation of the prostate in trans-rectal ultrasound \cite{samei2021automatic}. 
In these applications, SSMs typically need to be deformably registered with an organ instance in the image space \cite{raju2022deep}. 
However, SSMs have also been employed in a broader context of registration tasks, such as in percutaneous ultrasound \cite{chan2003integration}, as a regularization tool in radiation planning \cite{berendsen2013free}, or in the correction of cardiac slices \cite{banerjee2021optimised}.

Medical image atlases provide an integrated representation of shape and appearance, offering additional features beneficial for registration.
However, building and utilizing an image atlas can be computationally intensive, and the quality heavily depends on accurate image registration, which can incur significant manual labor \cite{fu2020deep}.
Automated methods for deformable registration have been extensively explored \cite{fu2020deep}. 
However, less-than-perfect registration results can yield voxels that are not representative of any actual human anatomy, adversely affecting downstream applications.
This work focuses on SSMs in the form of point distribution models (PDMs) obtained through unsupervised correspondence estimation \cite{bastian2023s3m}, which are preferred in some applications as a generalizable and lightweight statistical organ representation \cite{adams2023learning}.

\subsection{Multi-modal Registration}

Multi-modal registration has been established as a cornerstone for surgical navigation as well as pre-operative and general acquisition planning \cite{hennerspergerMRIBasedAutonomousRobotic2017}.
Several data modalities have been proposed for navigational support, including multi-template medical atlases~\cite{guerreiro2017evaluation}, or using MRI for US acquisition planning~\cite{hennerspergerMRIBasedAutonomousRobotic2017}.
Deep-learning-based methods have recently proven useful for multi-modal registration due to their robustness to initialization and accuracy \cite{markovaGlobalMultimodal2D2022a}.
Markova et. al. propose to learn dense features from MRI and US modalities, which are combined in a matching module using a confidence threshold and processed with RANSAC to retrieve the pose.
Alternatively, learning modality-invariant features can be achieved by sampling triplets and driving together similar latent descriptions of two deep neural networks through a contrastive or triplet loss \cite{feng2D3DMatchNetLearningMatch2019}.

While multi-modal slice-to-volume registration is of high interest to the medical community \cite{ferrante2017slice,markovaGlobalMultimodal2D2022a}, the registration of image slices to statistical shape representations has not been explored extensively.
Ghanavati et al. first proposed registering US slices to an image atlas generated by deformable registration of CT images \cite{ghanavati2010multi,ghanavati2011phantom}.
More recently, Yeung et al. proposed localizing slices within an atlas of the fetal brain \cite{yeungAdaptive3DLocalization2022}.
In contrast to these methods, we propose to localize slices within a PDM.
This incurs significant challenges, as PDMs lack image intensity features and, unlike image atlases, only contain sparse and noisy surface samples.
The localization of slices to a PDM could overcome the deficits image atlases suffer due to registration inaccuracies, as surface correspondence estimation is generally less ill-posed than the deformable registration of dense volumes.
Unlike previous works, we therefore propose the task of \textit{slice-to-shape} registration. 
We use a triplet-learning loss to construct a common cross-domain embedding space between US images and organ contours, enabling image localization with respect to a 3D statistical geometry.

\section{Method}
\label{sec:method}
In the following, we propose a deep learning framework to register a US image slice to a 3D shape of the same organ. 
We first leverage self-supervised contrastive learning to learn a common latent feature space between patches extracted from compounded 3D US data and patches representing local portions of the 3D organ surface. 
We then sample patches from both modalities during inference, establishing an optimal matching via distance in the latent space.
The slice location can then be approximated based on these matches through several iterations of a refinement algorithm, eliminating false-positive matches and narrowing down the region on the surface where the slice is likely located.

We use a discretized signed distance field (SDF) to represent the 3D thyroid surface.
The SDF is constructed from one of three possible sources, depending on the task: 
\begin{itemize}
    \item a patient-specific mesh created from the 3D ultrasound segmentation labels (registering a US slice to the 3D SDF of the same patient)
    \item the SSM's mean shape
    \item a shape sampled from the SSM distribution
\end{itemize}

For the SSM, we construct a point distribution model (PDM) to use as a registration target for US patch features.
Points are first brought into correspondence using S3M \cite{bastian2023s3m} to form a matrix $X\in\mathbf{R}^{n\times d}$, with $d$ point coordinates for each of the $n$ samples.
One can then form the mean shape $\bar{X}\in\mathbf{R}^d$ and covariance matrix $S\in\mathbf{R}^{d\times d}$ over the $n$ samples.
Since S has rank $n-1$, the matrix has $n-1$ eigenvectors $v_j$ with eigenvalues $\lambda_j$.
Considering the sum
\begin{align}\label{pdm_eq}
    s = \bar{X} + \sum_{j=1}^{n-1}\alpha_j\lambda_j v_j,\qquad \alpha_j\sim\mathcal{N}(0,1)
\end{align}
then $s\sim\mathcal{N}\left(\bar{X},S\right)$, which describes the desired distribution of the SSM. Next, we use the correspondences $X$ to register the US scan with the SDF voxel grid.

\begin{figure}[t]
    \centering
    \includegraphics[width=\textwidth]{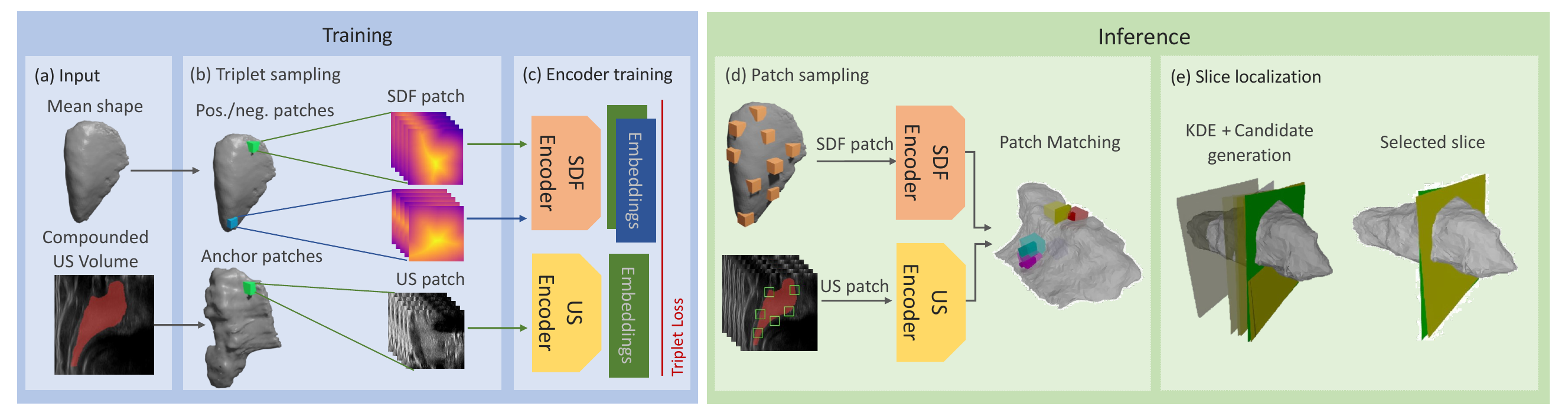}
    \caption{Our framework for US localization within a PDM. The input comprises US data and a 3D shape model (a). Anchor, positive, and negative patches are sampled from US and SDF modalities (b), after which we learn a joint embedding across modalities with a triplet loss (c). During inference, sampled patches are encoded and matched based on their similarity in the embedding space (d). We employ Kernel Density Estimation (KDE) and an iterative refinement to localize the position of a candidate slice on the 3D shape model (e).}
    \label{fig:pipeline}
\end{figure}

We proceed by learning a joint embedding space between the compounded US images and the SDF for each of the three tasks.
During inference, an unseen US scan can be localized within the learned embedding space of the SDF by comparing patches in the embedding space of both networks. 
For inference, we use axis-aligned slices extracted from the compounded volume instead of raw B-mode US slices and make the simplifying assumption that slices have a certain thickness in the longitudinal (z-axis) direction.
Figure \ref{fig:pipeline} depicts the proposed method.

\subsection{Encoder training}
Two 3D CNN encoders \cite{cciccek20163dunet} are used to encode cube-shaped patches from each modality into a common embedding space.
These are trained with a weighted soft-margin triplet loss \cite{feng2D3DMatchNetLearningMatch2019} to ensure that geometrically corresponding patches are mapped to similar regions in the embedding space. 
Our triplets consist of anchor patches sampled from the US data and corresponding positive/negative patches sampled from the SDF grid. 
All patches are sampled near the thyroid surface. 
Anchor patches from the US data and positive/negative patches from the SDF are then fed into the two respective encoders.
Given the embeddings $e_0$ of a US anchor patch and $e_+, e_-$ of the corresponding positive/negative SDF patches, respectively, as well as a hyperparameter $\alpha$, the weighted soft-margin loss is \begin{equation}
    L(e_0, e_+, e_-) = \log(1 + \exp(\alpha \left( ||e_0 - e_+|| - ||e_0 - e_-|| \right)))
\end{equation}
To generate an approximately uniform distribution of samples across the organ's surface, we first use farthest point sampling~\cite{moenningFastMarchingFarthest2003} on the vertices of the scan-specific segmentation label mesh (whose coordinates agree with the US image space).
A positive SDF patch is then defined as being in the same position as the US anchor patch: 
For the SSM samples, this requires transferring the coordinates to the semantically corresponding location in the SDF grid, which we achieve by using the learned point-to-point correspondences of S3M \cite{bastian2023s3m}.
For each positive sample, a negative SDF patch is sampled uniformly from the SDF mesh vertices among a percentile (e.g., 50\%) of vertices furthest away from the positive patch center.

\subsection{Slice localization}
During inference, we proceed by regressing a US image slice based on similar patches across the two embedding spaces.
We first sample patch centers from the two modalities to localize a US image in the SDF representation. 
SDF patches are sampled via farthest point sampling of the mesh vertices.
We sample patches from the US slice near the thyroid surface using the available ground-truth segmentation labels. In practice, these could be obtained by a real-time segmentation network as in \cite{kronkeTracked3DUltrasound2022}.

Next, patches are fed through the respective encoders to obtain patch embeddings, to which we apply the Hungarian matching algorithm using the Euclidean distance to establish cross-modality matches between similar patches in the embedding space.
From these matches, we then estimate the slice location using the Procrustes algorithm~\cite{schonemannGeneralizedSolutionOrthogonal1966}. 

To reduce the impact of false-positive matches, we run an iterative search, narrowing down the search space based on the matches of the previous iteration. 
To this end, we employ a kernel density estimation along the longitudinal axis of the thyroid, and several local maxima are taken as candidate locations.
For each candidate location, the cross-modality matching algorithm is repeated using SDF patches only sampled around this neighborhood, and a slice candidate is generated via the Procrustes algorithm.
The final prediction is the candidate slice with the lowest Procrustes loss.

In detail, the iterative algorithm works as follows: the coordinates of the SDF patches identified as matches are projected to the longitudinal axis ($z$ axis in our coordinate system), and a kernel density estimation (KDE) that fits a Gaussian mixture density to the projected coordinates is computed. 
The $m_{\text{KDE}}$ largest local maxima of the density are found and taken as candidate locations. 
For each candidate $z$-coordinate $z_j$, a restricted mesh is computed that only contains vertices with a $z$ coordinate in the range $[z_j \pm w_{\text{restr}}]$. 
This restricted mesh is used to sample SDF patches near the mesh surface for this candidate in the next iteration. 
The process is repeated until $m_{\text{step}}$ restriction steps have been performed. 
The slice transformation with the lowest Procrustes loss is output as the algorithm's prediction.
The hyperparameters $m_{\text{KDE}}$, $w_{\text{restr}}$ and $m_{\text{step}}$ as well as the number of sampled SDF and US patches, are tuned on a subset of the data.

In the above, we restrict ourselves to slices parallel to the axial image plane, a restriction imposed by the kernel density estimation algorithm we require for outlier elimination. 
However, this is the same direction in which the thyroid is typically scanned during diagnosis and should, therefore, not be limiting in practice.

\section{Experiments}
\noindent\textbf{Thyroid Dataset.}
We evaluate our method on a publically available dataset of freehand US scans of healthy thyroids acquired from volunteers aged 24 - 39 years~\cite{kronkeTracked3DUltrasound2022}. 
Each US sweep is compounded to a 3D resolution of 0.12 x 0.12 x 0.12 millimeters (mm). 
We use the right thyroid lobes of a subset of 16 patients for which ground truth segmentation maps are available.
Although the US slices were labeled under the supervision of radiologists, the segmentation boundaries contain significant noise, making the correspondence and matching tasks challenging.
Figure \ref{fig:example_thyroids} (a) depicts various right thyroid lobes from the patient dataset.
The thyroid dataset has high variance in the segmentation contours of a given sample and in the overall anatomical size and morphology between samples~\cite{kronkeTracked3DUltrasound2022}.

\subsection{Multi-modal Registration}

We evaluate the proposed multi-modal registration method by matching US slices to three types of 3D surface representations (c.f. Figure \ref{fig:example_thyroids}): patient-specific 3D labels, the mean shape of an SSM, and shapes sampled from an SSM, as described in section \ref{sec:method}.

All experiments in this section are carried out with a 4-fold cross-validation over the 16 samples. 
The slice localization accuracy is also tested with two input voxel patch sizes at the 0.12 mm compounding resolution: (32, 32, 32) and (64, 64, 8). The first two dimensions correspond to the x-y axes of the axial plane, while the longitudinal axis (z coordinate) projects into the axial image plane. The longitudinal axis is the direction along which all US sweeps are acquired.
The patches protrude 3.84 mm and 0.96 mm into the longitudinal axis, respectively.

The following experiments are conducted to evaluate the proposed method.
For US image to patient-SDF registration, we learn to match US slices to the 3D SDF representation of the same US acquisition.
We also ablate over two training strategies to train an encoder that enables matching from the US slices to the shape model. 
We train the 3D SDF encoder with patches from the mean shape or shapes sampled from the SSM, with $\alpha_j \sim \mathcal N(0, 0.5)$ according to Equation \ref{pdm_eq}.
The latter strategy could be considered a form of data augmentation.

To evaluate the generalization capability of our model to the slice matching task, we generate 50 slices evenly spaced along the longitudinal axis (parallel to the axial image plane) for each validation thyroid lobe.
The models trained to match US image patches to SSM samples are all evaluated for their slice prediction capabilities on the mean shape. 
We report the following metrics for the proposed matching method: the translational error and absolute rotational error between the predicted and ground truth slice and the percentage of predictions with translational error less than 10\% and 15\% of the longitudinal-axis length of the mean thyroid lobe.
The mean shapes generated from the train set of each cross-validation fold have a length of $39.5 \pm 0.655$ mm.

\section{Results \& Discussion}

\begin{table}[!htb]
    \setlength{\tabcolsep}{7pt}
    \renewcommand{\arraystretch}{1.2}
    \centering
    \caption{Results for the US slice registration to either an SDF representation of the patient anatomy or the mean shape from our SSM for two encoder patch sizes. We evaluate the translational (mm) and rotational error (in absolute degrees) of the predicted slice to the ground truth US slice and the percentage of slices within $10\%$ and $15\%$ distance along the z-axis of the mean shape in mm. For all experiments, the registration target is either the SDF of an individual patient (Patient SDF) or the mean shape of an SSM (Mean Shape). During training, we ablate over learning to match features to a Patient SDF, Mean Shape, or SSM samples with $\alpha \sim N(0, 0.5)$ (see Section \ref{sec:method}).}
    \resizebox{0.99\textwidth}{!}{
    \begin{tabular}{c c | c c c c}
    \toprule
    Patch Dimension & Train Source / Reg. Target & Trans. error (mm) & Rot. error ($^\circ$) & $10\%$ thresh. & $15\%$ thresh. \\
    \midrule
    \multirow{4}{*}{(32, 32, 32) \vspace{0.5cm}} 
        & Patient SDF & 1.82 $\pm$ 0.10 & 2.32 $\pm$ 0.15 & 90.97\% $\pm$ 1.02 & 95.71\% $\pm$ 0.77 \\
        & Mean Shape & 4.60 $\pm$ 0.42 & 2.39 $\pm$ 0.25 & 51.16\% $\pm$ 7.28 & 75.59\% $\pm$ 4.27 \\
        & SSM / Mean Shape & 5.08 $\pm$ 0.40 & 2.64 $\pm$ 0.48 & 42.38\% $\pm$ 9.70 & 68.75\% $\pm$ 6.13 \\
    \midrule
    \multirow{4}{*}{(64, 64, 8) \vspace{0.5cm}}
        & SDF Patient & 1.21 $\pm$ 0.08 & 2.27 $\pm$ 0.08 & 96.47\% $\pm$ 0.46 & 98.75\% $\pm$ 0.49 \\
        & Mean Shape & 4.95 $\pm$ 0.41 & 3.90 $\pm$ 1.52 & 44.38\% $\pm$ 10.33 & 71.50\% $\pm$ 3.51 \\
        & SSM / Mean Shape & 5.38 $\pm$ 0.54 & 4.12 $\pm$ 1.05 & 40.38\% $\pm$ 9.04 & 68.38\% $\pm$ 5.91 \\
    \bottomrule
    \end{tabular}
    }
    \label{tab:registration_crossval}
\end{table}

\noindent\textbf{Multi-modal Registration.} Table \ref{tab:registration_crossval} depicts the results of our multi-modal registration model.
The model can accurately register US slices to the 3D SDF contours from the same patient anatomy for both input embedding patch sizes. 
The accuracy obtained for this registration is as low as 1.21 $\pm$ 0.08mm translational error and 2.27$^\circ$ rotational error, with 96.47\% of slices predicted within 10\% of the thyroid lobe length for the (64, 64, 8) patch shape.
These results demonstrate that it is possible to match US acquisition slices to a patient-specific topological thyroid representation with an accuracy sufficient for acquisition and surgical planning \cite{chengUse3DUltrasound2022}.

The best overall slice matching performance on registration to the mean shape is achieved with a patch shape of (32, 32, 32), yielding an average translational error of 4.60 mm and a rotational error of 2.39$^\circ$, with 75.59\% of predicted slices located within 15\% of the ground truth slice, and 51.16\% within 10\%.

Both of these slice-matching methods have noteworthy practical ramifications in a clinical setting.
For example, if a nodule detected in a patient's thyroid is suspected to be malignant, the patient will be recommended for biopsy or follow-up screenings.
Furthermore, our methodology generalizes to the more general mean shape representation, albeit with lower registration accuracy.
Localizing US slices in this manner could enable acquisition planning with respect to an SSM without additional patient-specific acquisitions.
SSMs can be easily shared and deployed in acquisition systems. 
They can be represented compactly and do not contain possibly identifying information.
This could ease the accessibility of such a system for clinics that do not have the resources to curate large medical atlases.

\subsection{Limitations and Future Work} 
To sample image patches corresponding to the SDF's surface, our method requires a segmentation contour of the thyroid lobe during training and inference.
In future work, this could be mitigated during inference by learning to segregate features on the boundary of the organ from other parts of the image during training.
Furthermore, our US slice to SSM matching considers only the mean shape and samples from the distribution during inference.
Using samples from the SSM as data augmentation proves inferior to matching directly to the PDM mean shape.
However, future works could explore learning to encode the entire SSM distribution for patch correspondence and general slice localization, as this could increase localization accuracy.
\section{Conclusion}

This work presents an automated method for US slice localization to aid in surgical and acquisition planning. 
By formulating the localization problem as a 2D-to-3D registration to a 3D SDF, the proposed method localizes 2D US slices within two different geometric representations of the patient's anatomy.
We demonstrate that our unsupervised correspondence method is robust to the heterogenous and noisy thyroid topology across a set of individuals.
Furthermore, we propose a pipeline that enables registration of a US slice to not only the surface of the patient anatomy but also a more general statistical representation across a population.
Consistent localization of 2D US slices without a previous acquisition could enable several applications, including improved automated robotic scanning, sonographic acquisition planning, or guidance for hands-on US or anatomical training.
Perhaps a glimpse into the complex thyroid anatomy in the form of a single US image can yield more insight than previously realized.
We are confident this work will advance research in automated thyroid scanning and diagnosis, which has the potential to improve the quality of life of millions suffering from thyroid disorders worldwide.\\

\noindent\textbf{Acknowledgements and Disclosure.}
The thyroid dataset used for all experiments is publicly available. Vincent Bürgin is supported by the DAAD program Konrad Zuse Schools of Excellence in Artificial Intelligence, sponsored by the Federal Ministry of Education and Research. The authors declare no conflicts of interest.

\bibliographystyle{splncs04}
\bibliography{references}

\begin{thebibliography}{10}
\providecommand{\url}[1]{\texttt{#1}}
\providecommand{\urlprefix}{URL }
\providecommand{\doi}[1]{https://doi.org/#1}

\bibitem{adams2023learning}
Adams, J., Khan, N., Morris, A., Elhabian, S.: Learning spatiotemporal
  statistical shape models for non-linear dynamic anatomies. Frontiers in
  Bioengineering and Biotechnology  \textbf{11},  1086234 (2023)

\bibitem{azizi3DUltrasoundThyroid2021}
Azizi, G., Faust, K., Ogden, L., Been, L., Mayo, M.L., Piper, K., Malchoff, C.:
  3-{{D Ultrasound}} and {{Thyroid Cancer Diagnosis}}: {{A Prospective Study}}.
  Ultrasound in Medicine \& Biology  \textbf{47}(5) (2021)

\bibitem{banerjee2021optimised}
Banerjee, A., Zacur, E., Choudhury, R.P., Grau, V.: Optimised misalignment
  correction from cine mr slices using statistical shape model. In: Annual
  Conference on Medical Image Understanding and Analysis. pp. 201--209.
  Springer (2021)

\bibitem{bastian2023s3m}
Bastian, L., Bauman, A., Hoppe, E., B{\"u}rgin, V., Kim, H.Y., Saleh, M.,
  Busam, B., Navab, N.: {S3M}: Scalable statistical shape modeling through
  unsupervised correspondences. arXiv preprint arXiv:2304.07515  (2023)

\bibitem{berendsen2013free}
Berendsen, F.F., Van Der~Heide, U.A., Langerak, T.R., Kotte, A.N., Pluim, J.P.:
  Free-form image registration regularized by a statistical shape model:
  application to organ segmentation in cervical mr. Computer Vision and Image
  Understanding  \textbf{117}(9),  1119--1127 (2013)

\bibitem{chai2022successful}
Chai, H.h., Ye, R.z., Xiong, L.f., Xu, Z.n., Chen, X., Xu, L.j., Hu, X., Jiang,
  L.f., Peng, C.z.: Successful use of a 5g-based robot-assisted remote
  ultrasound system in a care center for disabled patients in rural china.
  Frontiers in Public Health  \textbf{10},  915071 (2022)

\bibitem{chan2003integration}
Chan, C.S., Edwards, P.J., Hawkes, D.J.: Integration of ultrasound-based
  registration with statistical shape models for computer-assisted orthopedic
  surgery. In: Medical Imaging 2003: Image Processing. vol.~5032, pp. 414--424.
  SPIE (2003)

\bibitem{chengUse3DUltrasound2022}
Cheng, A., Lee, J.W.K., Ngiam, K.Y.: Use of {{3D}} ultrasound to characterise
  temporal changes in thyroid nodules: An in vitro study. Journal of Ultrasound
   (2022)

\bibitem{cciccek20163dunet}
{\c{C}}i{\c{c}}ek, O., Abdulkadir, A., Lienkamp, S.S., Brox, T., Ronneberger,
  O.: {3D} {U}-{N}et: learning dense volumetric segmentation from sparse
  annotation. In: MICCAI 2016

\bibitem{ellingsenRobustDeformableImage2010}
Ellingsen, L.M., Chintalapani, G., Taylor, R.H., Prince, J.L.: Robust
  deformable image registration using prior shape information for atlas to
  patient registration  \textbf{34}(1) (2010)

\bibitem{feng2D3DMatchNetLearningMatch2019}
Feng, M., Hu, S., Ang, M.H., Lee, G.H.: 2d3d-matchnet: Learning to match
  keypoints across 2d image and 3d point cloud. In: ICRA 2019

\bibitem{ferrante2017slice}
Ferrante, E., Paragios, N.: Slice-to-volume medical image registration: A
  survey. Medical image analysis  \textbf{39},  101--123 (2017)

\bibitem{fu2020deep}
Fu, Y., Lei, Y., Wang, T., Curran, W.J., Liu, T., Yang, X.: Deep learning in
  medical image registration: a review. Physics in Medicine \& Biology
  \textbf{65}(20),  20TR01 (2020)

\bibitem{ghanavati2011phantom}
Ghanavati, S., Mousavi, P., Fichtinger, G., Abolmaesumi, P.: Phantom validation
  for ultrasound to statistical shape model registration of human pelvis. In:
  Medical Imaging 2011: Visualization, Image-Guided Procedures, and Modeling.
  vol.~7964, pp. 855--862. SPIE (2011)

\bibitem{ghanavati2010multi}
Ghanavati, S., Mousavi, P., Fichtinger, G., Foroughi, P., Abolmaesumi, P.:
  Multi-slice to volume registration of ultrasound data to a statistical atlas
  of human pelvis. In: Medical Imaging 2010: Visualization, Image-Guided
  Procedures, and Modeling. vol.~7625, pp. 213--222. SPIE (2010)

\bibitem{grassi2021statistical}
Grassi, L., V{\"a}{\"a}n{\"a}nen, S.P., Isaksson, H.: Statistical shape and
  appearance models: Development towards improved osteoporosis care. Current
  Osteoporosis Reports  \textbf{19},  676--687 (2021)

\bibitem{guerreiro2017evaluation}
Guerreiro, F., Burgos, N., Dunlop, A., Wong, K., Petkar, I., Nutting, C.,
  Harrington, K., Bhide, S., Newbold, K., Dearnaley, D., et~al.: Evaluation of
  a multi-atlas ct synthesis approach for mri-only radiotherapy treatment
  planning. Physica Medica  (2017)

\bibitem{haugen2015AmericanThyroid2016}
Haugen, B.R., Alexander, E.K., Bible, K.C., Doherty, G.M., Mandel, S.J.,
  et~al.: 2015 {{American Thyroid Association Management Guidelines}} for
  {{Adult Patients}} with {{Thyroid Nodules}} and {{Differentiated Thyroid
  Cancer}}: {{The American Thyroid Association Guidelines Task Force}} on
  {{Thyroid Nodules}} and {{Differentiated Thyroid Cancer}}. Thyroid
  \textbf{26}(1) (2016)

\bibitem{heimannStatisticalShapeModels2009}
Heimann, T., Meinzer, H.P.: Statistical shape models for {{3D}} medical image
  segmentation: {{A}} review  \textbf{13}(4),  543--563 (2009)

\bibitem{hennerspergerMRIBasedAutonomousRobotic2017}
Hennersperger, C., Fuerst, B., Virga, S., Zettinig, O., Frisch, B., Neff, T.,
  Navab, N.: Towards {{MRI-Based Autonomous Robotic US Acquisitions}}: {{A
  First Feasibility Study}}  \textbf{36}(2),  538--548 (2017)

\bibitem{hu2022deep}
Hu, X., Chen, X., Liu, Y., Chen, E.Z., Chen, T., Sun, S.: Deep statistic shape
  model for myocardium segmentation. arXiv preprint arXiv:2207.10607  (2022)

\bibitem{kronkeTracked3DUltrasound2022}
Kr{\"o}nke, M., Eilers, C., Dimova, D., K{\"o}hler, M., Buschner, G., et~al.:
  Tracked 3d ultrasound and deep neural network-based thyroid segmentation
  reduce interobserver variability in thyroid volumetry. Plos one
  \textbf{17}(7) (2022)

\bibitem{ludke2022landmark}
L{\"u}dke, D., Amiranashvili, T., Ambellan, F., Ezhov, I., Menze, B.H., Zachow,
  S.: Landmark-free statistical shape modeling via neural flow deformations.
  In: International Conference on Medical Image Computing and Computer-Assisted
  Intervention. pp. 453--463. Springer (2022)

\bibitem{markovaGlobalMultimodal2D2022a}
Markova, V., Ronchetti, M., Wein, W., Zettinig, O., Prevost, R.: Global
  {{Multi-modal 2D}}/{{3D Registration}} via {{Local Descriptors Learning}}.
  In: MICCAI 2022

\bibitem{moenningFastMarchingFarthest2003}
Moenning, C., Dodgson, N.A.: Fast {{Marching}} farthest point sampling.
  Eurographics 2003 - Posters  (2003)

\bibitem{naceri2022tactile}
Naceri, A., Elsner, J., Tr{\"o}binger, M., Sadeghian, H., Johannsmeier, L.,
  Voigt, F., Chen, X., Macari, D., J{\"a}hne, C., Berlet, M., et~al.: Tactile
  robotic telemedicine for safe remote diagnostics in times of corona: System
  design, feasibility and usability study. IEEE Robotics and Automation Letters
   \textbf{7}(4),  10296--10303 (2022)

\bibitem{raju2022deep}
Raju, A., Miao, S., Jin, D., Lu, L., Huang, J., Harrison, A.P.: Deep implicit
  statistical shape models for 3d medical image delineation. In: proceedings of
  the AAAI conference on artificial intelligence. vol.~36, pp. 2135--2143
  (2022)

\bibitem{samei2021automatic}
Samei, G., Karimi, D., Kesch, C., Salcudean, S.: Automatic segmentation of the
  prostate on 3d trans-rectal ultrasound images using statistical shape models
  and convolutional neural networks. arXiv preprint arXiv:2106.09662  (2021)

\bibitem{schonemannGeneralizedSolutionOrthogonal1966}
Sch{\"o}nemann, P.H.: A generalized solution of the orthogonal procrustes
  problem. Psychometrika  \textbf{31}(1),  1--10 (1966)

\bibitem{songCrossModalAttentionMRI2021}
Song, X., Guo, H., Xu, X., Chao, H., Xu, S., Turkbey, B., Wood, B.J., Wang, G.,
  Yan, P.: Cross-{{Modal Attention}} for {{MRI}} and {{Ultrasound Volume
  Registration}}. In: MICCAI 2021

\bibitem{tang2019augmentation}
Tang, Z., Chen, K., Pan, M., Wang, M., Song, Z.: An augmentation strategy for
  medical image processing based on statistical shape model and 3d thin plate
  spline for deep learning. IEEE Access  \textbf{7},  133111--133121 (2019)

\bibitem{uzunova2022systematic}
Uzunova, H., Wilms, M., Forkert, N.D., Handels, H., Ehrhardt, J.: A systematic
  comparison of generative models for medical images. International Journal of
  Computer Assisted Radiology and Surgery  \textbf{17}(7),  1213--1224 (2022)

\bibitem{yeungAdaptive3DLocalization2022}
Yeung, P.H., Aliasi, M., Haak, M., Xie, W., Namburete, A.I.: Adaptive {{3D
  Localization}} of {{2D Freehand Ultrasound Brain Images}}. In: MICCAI 2022

\bibitem{zhangClinicalApplication5Gbased2022}
Zhang, Y.Q., Yin, H.H., He, T., Guo, L.H., Zhao, C.K., Xu, H.X.: Clinical
  application of a {{5G-based}} telerobotic ultrasound system for thyroid
  examination on a rural island: A prospective study  \textbf{76}(3),  620--634

\end{thebibliography}

\end{document}